%% file: main.tex
\documentclass[runningheads]{llncs}
\usepackage{algorithm}
\usepackage{algpseudocode}
\usepackage{amsfonts, amssymb}
\usepackage{color, soul, xcolor}
\usepackage[utf8]{inputenc}
\usepackage{listings}
\usepackage{physics}

\usepackage[english]{babel}
\usepackage[nocompress]{cite}
\usepackage[inline]{enumitem}
\usepackage{float}
\usepackage{graphicx}
\usepackage{hyperref}
\usepackage[parfill]{parskip}
\usepackage{subcaption}

\def\Schrodinger{Schr{ö}dinger}
\def\eg{{\em e.g., }}
\def\ie{{\em i.e., }}
\def\Rn{{\mathbb{R}^n}}
\def\Cn{{\mathbb{C}^n}}
\def\dBdt{{\frac{\dd B}{\dd t}}}
\def\funcfam{conjugate-flattening}
\def\Funcfam{Conjugate-flattening}
\def\boldalpha{\boldsymbol{\alpha}}
\def\boldbeta{\boldsymbol{\beta}}
\def\iu{{\mathrm{i}}}

\newcommand{\init}[1]{#1_0}
\newcommand{\unsafe}[1]{#1_u}
\newcommand{\prob}[1]{\abs{#1}^2}
\newcommand{\replaceu}[1]{\left. #1\right\vert_{u=\overline{z}}}

\begin{document}

\title{Verification of Quantum Systems using Barrier Certificates}

\author{Marco Lewis\inst{1}\thanks{Corresponding email: m.j.lewis2@newcastle.ac.uk}
\orcidID{0000-0002-4893-7658}
\and
Paolo Zuliani\inst{1,2}\thanks{Currently at Universit\`{a} di Roma; work predominately done at Newcastle University.}
\orcidID{0000-0001-6033-5919}
\and
Sadegh Soudjani\inst{1,3}\orcidID{0000-0003-1922-6678}
}
\authorrunning{M. Lewis, P. Zuliani, S. Soudjani}

\institute{Newcastle University, Newcastle upon Tyne, UK
\and
Universit\`{a} di Roma ``La Sapienza'', Rome, Italy
\and
Max Planck Institute for Software Systems, Germany
}

\maketitle

\begin{abstract}
Various techniques have been used in recent years for verifying quantum computers, that is, for determining whether a quantum computer/system satisfies a given formal specification of correctness.
Barrier certificates are a recent novel concept developed for verifying properties of dynamical systems.
In this article, we investigate the usage of barrier certificates as a means for verifying behaviours of quantum systems.
To do this, we extend the notion of barrier certificates from real to complex variables.
We then develop a computational technique based on linear programming to automatically generate polynomial barrier certificates with complex variables taking real values.
Finally, we apply our technique to several simple quantum systems to demonstrate their usage.

% Keyword
\end{abstract}

\section{Introduction}
\input{Sections/intro}

\section{Background}
\label{sec:bg}
\input{Sections/barrier_certs}

\section{Complex-valued Barrier Certificates}
\label{sec:cbc}
\input{Sections/complex_bcs}

\section{Generating Satisfiable Barrier Certificates for Quantum Systems}
\label{sec:comp}
\input{Sections/computation}

% Example using the \Schrodinger{} Equation
\section{Application to Quantum Systems}
\label{sec:ex}
\input{Sections/quantum}

\section{Conclusions \label{sec:conc}}
In this paper, we extended the theory of barrier certificates to handle complex variables and demonstrated that barrier certificates can be extended to use complex variables.
We then showed how one can automatically generate simple complex-valued barrier certificates using polynomial functions and linear programming techniques.
Finally, we explored the application of the developed techniques by investigating properties of time-independent quantum systems.

There are numerous directions for this research to take.
In particular, one can consider (quantum) systems that are time-dependent, have a control component or are discrete-time, \ie quantum circuits.
Data-driven approaches for generating barrier certificates based on measurements of a quantum system can also be considered.
A final challenge to consider is how to verify large quantum systems.
Techniques, such as Trotterization, allow Hamiltonians to be simulated either by simpler Hamiltonians of the same size or of lower dimension.
How barrier certificates can ensure safety of such systems is a route to explore.

\section*{Acknowledgements}
M.Lewis is supported by the UK EPSRC (project reference EP/T517914/1). The work of S. Soudjani is supported by the following grants: EPSRC EP/V043676/1, EIC 101070802, and ERC 101089047.

\paragraph{Data availability.}
The public repository with an implementation of the algorithm from Section~\ref{sec:comp} and case studies from Section~\ref{sec:ex} is available on GitHub: \url{https://github.com/marco-lewis/quantum-barrier-certificates}.

\bibliographystyle{splncs04}
\bibliography{ref}

\appendix
\input{Sections/appendix}

\end{document}

%% file: Sections/intro.tex
Quantum computers are powerful devices that allow certain problems to be solved faster than classical computers.
The research area focusing on the formal verification of quantum devices and software has witnessed the extension of verification techniques from classical systems~\cite{FVBook,modelchecking} to the quantum realm.
Classical techniques that have been used include
theorem provers \cite{SQIR,QHL},
Binary Decision Diagrams \cite{Zulehner2017,Burgholzer2021},
SMT solvers \cite{QBricks,Giallar} and
other tools \cite{ZXcalc,Honarvar2020}.

Quantum systems evolve according to the \Schrodinger{} equation from some initial state.
However, the initial state may not be known completely in advance.
One can prepare a quantum system by making observations on the quantum objects, leaving the quantum system in a basis state, but this omits the global phase which is not necessarily known after measurement.
Further, the system could be disturbed through some external influence before it begins evolving.
This can slightly change the quantum state from the basis state to a state in superposition or possibly an entangled state.

By taking into account these uncertain factors, a set of possible initial states from which the system evolves can be constructed.
From this initial set, we can see if the system evolves according to some specified behaviour such as reaching or avoiding a particular set of states.
As an example, consider a single qubit system that evolves according to a Hamiltonian $\hat{H}$ implementing the controlled-NOT operation.
Through measurement and factoring in for noise, we know the system starts close to $\ket{10}$.
The controlled-NOT operation keeps the first qubit value the same and so we want to verify that, as the system evolves via $\hat{H}$, the quantum state does not evolve close to $\ket{00}$ or $\ket{01}$.

The main purpose of this work is to study the application of a technique called \emph{barrier certificates}, used for verifying properties of classical dynamical systems, to check properties of quantum systems similar to the one mentioned above.
The concept of barrier certificates has been developed and used in Control Theory to study the safety of dynamical systems from a given set of initial states on real domains \cite{Prajna2007}.
This technique can ensure that given a set of initial states from which the system can start and a set of unsafe states, the system will not enter the unsafe set.
This is achieved through separating the unsafe set from the initial set by finding a \emph{barrier}.

Barrier certificates can be defined for both deterministic and stochastic systems in discrete and continuous time \cite{lavaei2021automated,ames2019control}.
The concept has also been used for verification and synthesis against complicated logical requirements beyond safety and reachability \cite{Jagtap2020}.
The conditions under which a function is a barrier certificate can be automatically and efficiently checked using SMT solvers \cite{bak2018t}.
Such functions can also be found automatically using learning techniques even for non-trivial dynamical systems \cite{peruffo2021automated}.

Dynamical systems are naturally defined on real domains ($\mathbb{R}^n$).
To handle dynamical systems in complex domains ($\mathbb{C}^n$), one would need to decompose the system into its real and imaginary parts and use the techniques available for real systems.
This has two disadvantages, the first being that this doubles the number of variables being used for the analysis.
The second disadvantage is that the analysis may be easier to perform directly with complex variables than their real components.
As quantum systems use complex values, it is desirable to have a technique to perform the reachability analysis using complex variables.

In this paper, we explore the problem of safety verification in quantum systems by extending barrier certificates from real to complex domains.
Our extension is inspired by a technique developed by Fang and Sun~\cite{Fang2013}, who studied the stability of complex dynamical systems using Lyapunov functions (where the goal is to check if a system eventually stops moving).
Further, we provide an algorithm to generate barrier certificates for  quantum systems and use it to generate barriers for several examples.

%% file: Sections/barrier_certs.tex
\subsection{Safety Analysis}
We begin by introducing the problem of safety for dynamical systems with real state variables $x\in\Rn{}$.
More details can be found in \cite{Prajna2007}.
A continuous dynamical system is described by
\begin{equation*}
    \dot{x} = \dv{x}{t} = f(x),\quad f:\Rn{} \to \Rn{},
\end{equation*}
where the evolution of the system is restricted to $X \subseteq \Rn{}$ and $f$ is usually Lipschitz continuous to ensure existence and uniqueness of the differential equation solution.
The  set $\init{X} \subseteq X$ is the set of initial states and the unsafe set $\unsafe{X} \subseteq X$ is the set of values that the dynamics $x(t)$ should avoid.
These sets lead to the idea of safety for real continuous dynamical systems:
\begin{definition}[Safety]
\label{def:safety}
A system, $\dot{x} = f(x)$, evolving over $X \subseteq \Rn{}$ is considered safe if the system cannot reach the unsafe set, $X_u \subseteq X$, from the initial set, $X_0 \subseteq X$.
That is for all $t \in \mathbb{R}_+$ and $x(0) \in \init{X}$, then $x(t) \notin \unsafe{X}$.
\end{definition}

The safety problem is to determine if a given system is safe or not.
Numerous techniques have been developed to solve this problem~\cite{Franzle19}.
Barrier certificates are discussed in Section~\ref{sec:bc}.
Here, we describe two other common techniques.

\paragraph{Abstract Interpretation}
One way to perform reachability analysis of a system is to give an abstraction \cite{Cousot77,Cousot2001} of the system's evolution.
Given an initial abstraction that over-approximates the evolution of the system, the abstraction is improved based on false bugs.
False bugs are generated when the current abstraction enters the unsafe space but the actual system does not.
This method has been investigated for quantum programs in \cite{Yu21}, where the authors can verify programs using up to 300 qubits.

\paragraph{Backward and Forward Reachability}
A second approach is to start from the unsafe region and reverse the evolution of the system from there.
A system is considered unsafe if the reversed evolution enters the initial region.
This is backward reachability.
Conversely, forward reachability starts from the initial region and is considered safe if the reachable region does not enter the unsafe region.
Both backward and forward reachability are discussed in \cite{Mitchell07,SS2014precise,SS2015quantitative}.

\subsection{Barrier Certificates}
\label{sec:bc}

Barrier certificates~\cite{Prajna2007} are another technique used for safety analysis.
This technique attempts to divide the reachable region from the unsafe region by putting constraints on the initial and unsafe set, and on how the system evolves.
The benefit of barrier certificates over other techniques is that one does not need to compute the system's dynamics at all to guarantee safety, unlike in abstract interpretation and backward (or forward) reachability.

A barrier certificate is a differentiable function, $B: \Rn{} \to \mathbb{R}$, that determines safety through the properties that $B$ has.
Generally, a barrier certificate needs to meet the following conditions:
\begin{align}
    B(x) \leq 0 &, \forall x \in \init{X}
    \label{eq:initcond}\\
    B(x) > 0 &, \forall x \in \unsafe{X}
    \label{eq:unsafecond}\\
    x(0) \in \init{X} \implies B(x(t)) \leq 0 &, \forall t \in \mathbb{R}_+.
    \label{eq:barriercond}
\end{align}
Essentially, these conditions split the evolution space into a (over-approximate) reachable region  and an unsafe region, encapsulated by Conditions~\eqref{eq:initcond} and \eqref{eq:unsafecond} respectively.
These regions are separated by a ``barrier'', which is the contour along $B(x) = 0$.

Condition~\eqref{eq:barriercond} prevents the system evolving into the unreachable region and needs to be satisfied for the system to be safe.
However, Condition~\eqref{eq:barriercond} can be replaced with stronger conditions that are easier to check.
For example, the definition of one simple type of barrier certificate is given.
\begin{definition}[Convex Barrier Certificate]
\label{def:convex}
For a system $\dot{x} = f(x)$, $X \subseteq \Rn{}$, $X_0 \subseteq X$ and $X_u \subseteq X$, a function $B: \Rn{} \to \mathbb{R}$ that obeys the following conditions:
\begin{align}
    B(x) \leq 0 &, \forall x \in \init{X}\nonumber\\
    B(x) > 0 &, \forall x \in \unsafe{X}\nonumber \\
    \frac{\dd{B}}{\dd{x}} f(x) \leq 0 &, \forall x \in X, \label{eq:convexcond}
\end{align}
is a convex barrier certificate.
\end{definition}

Note that in Condition~\eqref{eq:convexcond}: $\dv{B}{x}\dv{x}{t} = \dv{B}{t}$.
This condition can be viewed as a constraint on the evolution of the barrier as the system evolves over time.

Now, if a system has a barrier certificate, then the system is safe.
We show the safety theorem for convex barrier certificates.
\begin{theorem}
\label{thm:real_convex_safe}
If a system, $\dot{x} = f(x)$, has a convex barrier certificate, $B: \Rn{} \to \mathbb{R}$, then the system is safe \cite{Prajna2007}.
\end{theorem}

\begin{figure}[t]
    \centering
    \includegraphics[width=.7\textwidth]{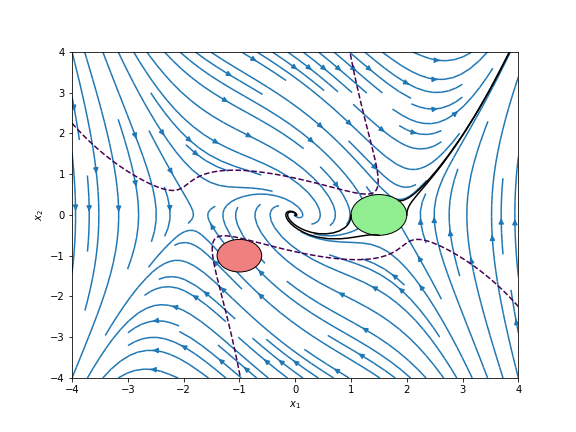}
    \caption{Example adapted from Section~{V-A} in \cite{Prajna2007}.
    The initial region is the green circle centred at $(1.5,0)$ and the system evolves according to the dynamical system given by differential equations ${\dot{x} = [x_2, -x_1 + \frac{1}{3}x_1^3 - x_2]}$.
    The unsafe region is the red circle centred at $(-1,-1)$ and is separated from the initial region by a barrier, the dashed purple line defined by $B(x) = 0$ where
    $B(x) = -13 + 7 x_1^2 + 16 x_2^2 - 6 x_1^2 x_2^2 -
    \frac{7}{6} x_1^4 - 3 x_1 x_2^3 + 12 x_1 x_2 - \frac{12}{3} x_1^3 x_2$.
    }
    \label{fig:dynamical_sys}
\end{figure}

Proofs of Theorem~\ref{thm:real_convex_safe} are standard and can be found in, \eg{}\cite{Prajna2007}.
The intuition behind the proof is that since the system starts in the negative region and the barrier can never increase, then the barrier can never enter the positive region.
Since the unsafe set is within the positive region of the barrier, this set can therefore never be reached.
Thus, the system cannot evolve into the unsafe set and so the system is safe.
Figure~\ref{fig:dynamical_sys} shows an example of a dynamical system with a barrier based on the convex condition.

\begin{remark}
The term ``convex'' is used for these barriers as the set of barrier certificates satisfying the conditions in Definition~\ref{def:convex} is convex.
In other words, if $B_1$ and $B_2$ are barrier certificates for a system, the function $\lambda B_1 + (1-\lambda) B_2$ is also a barrier certificate for any $\lambda\in[0,1]$.
See \cite{Prajna2007} or the proof of Proposition~\ref{prop:complexisconvex} in Appendix~\ref{app:complexisconvex} for (similar) details.
\end{remark}

There are a variety of different barrier certificates to choose from with different benefits, \eg{}the convex condition given is simple but may not work for complicated or nonlinear systems.
In comparison, the non-convex condition given in \cite{Prajna2007} changes Condition~\eqref{eq:convexcond} such that $\frac{\dd{B}}{\dd{x}} f(x) \leq 0; \forall x \in X, B(x) = 0$ (instead of $\forall x \in X$).
This is a weaker condition allowing for more functions to be a suitable barrier certificate.
However, a different computational method is required because the set of such barrier certificates is non-convex.
Each barrier certificate requires a different proof that if the system has a satisfying barrier certificate, then the system is safe.
It should be noted that Theorem~\ref{thm:real_convex_safe} only has a one way implication, a system does not necessarily have a barrier certificate even if it is safe.
In \cite{Wisniewski13}, the authors showed the converse holds for systems defined on a compact manifold and using convex barrier certificates.

%% file: Sections/complex_bcs.tex
Now we wish to extend the use of barrier certificates into a complex space ($\Cn{}$).
We use $\iu = \sqrt{-1}$ as the imaginary unit in the rest of the paper.
The complex dynamical systems considered are of the form
\begin{equation*}
\dot{z} = \dv{z}{t} = f(z),\quad f:\Cn{} \to \Cn{},
\end{equation*}
which evolves in $Z \subseteq \Cn{}$.
The initial and unsafe sets are defined in the usual way except now we have $\init{Z} \subseteq Z$ and $\unsafe{Z} \subseteq Z$, respectively. The notion of safety for this system is similar to Definition~\ref{def:safety}.
\begin{definition}[Safety]
A complex system, $\dot{z} = f(z)$, with $Z \subseteq \Cn{}$, $\init{Z} \subseteq Z$ and $\unsafe{Z} \subseteq Z$, is considered safe if for any $z(0) \in \init{Z}$, then $\forall t \in \mathbb{R}^+, z(t) \notin \unsafe{Z}$.
\end{definition}

Whilst it is easy to extend the safety problem and required definitions into the complex plane, extending the notion of barrier certificates requires particular attention.
Conditions \eqref{eq:initcond}, \eqref{eq:unsafecond} and \eqref{eq:barriercond} are changed respectively to
\begin{align}
    B(z) \leq 0 &, \forall z \in \init{Z} \label{eq:gencomplexinit}; \\
    B(z) > 0 &, \forall z \in \unsafe{Z} \label{eq:gencomplexunsafe}; \\
    z(0) \in \init{Z} \implies B(z(t)) \leq 0 &, \forall t \in \mathbb{R}_+. \label{eq:gencomplexsep}
\end{align}
Many barrier certificates use differential equations to achieve Condition~\eqref{eq:gencomplexsep}, which restricts the class of functions that can be used.
This is because differentiable complex functions must satisfy the Cauchy-Riemann equations.

For our purposes, we consider a holomorphic function, $g(z): \Cn{} \to \mathbb{C}$, to be a function whose partial derivatives, $\pdv{g(z)}{z_j}$, are holomorphic on $\mathbb{C}$, \ie{}they satisfy the Cauchy-Riemann equations (for several variables).
That is for $z_j = x_j + \iu y_j$ and $g(z) = g(x, y) = u(x,y) + \iu v(x,y)$, then
\begin{equation*}
    \begin{aligned}
        \pdv{u}{x_j} = \pdv{v}{y_j} & \quad & \pdv{u}{y_j} = - \pdv{v}{x_j}.
    \end{aligned}
\end{equation*}
Using an adapted technique developed by Fang and Sun \cite{Fang2013} allows us to reason about barrier certificates in the complex plane.
We begin by introducing a family of complex functions that are key to our technique.

\begin{definition}[\Funcfam{} function]
A function, $b: \Cn{} \cross \Cn{} \to \Cn{}$, is \funcfam{} if $\forall z \in \Cn{}, b(z, \overline{z}) \in \mathbb{R}$.
\end{definition}

\begin{definition}[Complex-valued barrier function]
A function, $B: \mathbb{C}^{n} \to \mathbb{R}$, is a complex-valued barrier function if $B(z) = b(z, \overline{z})$ where $b: \Cn{} \cross \Cn{} \to \Cn{}$ is a \funcfam{}, holomorphic function.
\end{definition}

Suppose now that we have a system that evolves over time, $z(t)$. To use the complex-valued barrier function, $B(z(t))$, for barrier certificates we require the differential of $B$ with respect to $t$.
Calculating this differential reveals that
\begin{equation}
\label{eq:complexdiff}
\begin{aligned}
    \dv{B(z(t))}{t} = \dv{b(z(t),\overline{z(t)})}{t}
    & = 
    \replaceu{\dv{b(z, u)}{z}}
    \dv{z}{t}
    + 
    \replaceu{\dv{b(z, u)}{u}} \overline{\dv{z}{t}}
    \\
    & = 
    \replaceu{\dv{b(z, u)}{z}} f(z) 
    + 
    \replaceu{\dv{{b(z, u)}}{u}} \overline{f(z)},
\end{aligned}
\end{equation}
where
$\dv{b(z,u)}{z} = \begin{bmatrix} \pdv{b(z,u)}{z_1}, \pdv{b(z,u)}{z_2}, \dots, \pdv{b(z,u)}{z_n} \end{bmatrix}$
is the gradient of $b(z,u)$ with respect to $z$ and the gradient is defined with respect to $u$ in a similar way.
Given Equation~\eqref{eq:complexdiff}, barrier certificates that include a differential condition can be extended into the complex domain quite naturally.
For example, the convex barrier certificate is extended to the complex domain.
\begin{definition}[Complex-valued Convex Barrier Certificate]
\label{def:complexconvex}
For a system $\dot{z} = f(z)$, $Z \subseteq \Cn{}$, $\init{Z} \subseteq Z$ and $\unsafe{Z} \subseteq Z$; a complex-valued barrier function $B: \Cn{} \to \mathbb{R}$, $B(z) = b(z, \overline{z})$, that obeys the following conditions,
\begin{align}
    b(z, \overline{z}) \leq 0 &, \forall z \in \init{Z} \label{eq:complexinit}\\
    b(z, \overline{z}) > 0 &, \forall z \in \unsafe{Z} \label{eq:complexunsafe} \\
    \replaceu{\dv{b(z, u)}{z}} f(z) 
    + 
    \replaceu{\dv{b(z, u)}{u}} \overline{f(z)} \leq 0 &, \forall z \in Z \label{eq:complexconvex},
\end{align}
is a complex-valued convex barrier certificate.
\end{definition}

With this definition, we can ensure the safety of complex dynamical systems:

\begin{theorem}
\label{thm:complex_convex_safe}
If a complex system, $\dot{z} = f(z)$, has a complex-valued convex barrier certificate, $B: \Cn{} \to \mathbb{R}$, then the system is safe.
\end{theorem}

\begin{proposition}
\label{prop:complexisconvex}
The set of complex-valued barrier certificates satisfying the conditions of Definition~\ref{thm:complex_convex_safe} is convex.
\end{proposition}

The proofs of these results are given in Appendix~\ref{app:complex_convex_safe} and \ref{app:complexisconvex} respectively.

%% file: Sections/computation.tex
We now describe how to compute a complex-valued barrier function.
Throughout, let $\dot{z} = f(z)$, $Z \subseteq \Cn{}$, $\init{Z} \subseteq Z$ and $\unsafe{Z} \subseteq Z$ be defined as before.
We introduce a general family of functions that will be used as ``templates'' for complex barrier certificates.
\begin{definition}
A $k$-degree polynomial function is a complex function, $b: \Cn{} \to \mathbb{C}$, such that
\begin{equation}
\label{eq:poly}
b(z_1, \dots, z_{n}) = \sum_{\boldalpha{} \in A_{n,k}} a_{\boldalpha{}} z^{\boldalpha{}}
\end{equation}
where $A_{n,k} := \{\boldalpha{} = (\alpha_1, \dots, \alpha_{n}) \subseteq \mathbb{N}^{n} : \sum_{j=1}^{n} \alpha_j \leq k \}$,
$a_{\boldalpha{}} \in \mathbb{C}$,
and $z^{\boldalpha{}} = \prod_{j=1}^n z_j^{\alpha_j}$.
\end{definition}

The family of $k$-degree polynomials are polynomial functions where no individual term of the polynomial can have a degree higher than $k$.
Note that $k$-degree polynomial functions are holomorphic.
Further, some $k$-degree polynomials are \funcfam{}.
For example, the 2-degree polynomial
$b(z_1, u_1) = z_1 u_1$
is \funcfam{} since $z\overline{z} = \prob{z}$, whereas the 1-degree polynomial
$b(z_1, u_1) = z_1$
is not.
Thus, a subset of this family of functions are suitable to be used for barrier certificates as complex-valued barrier functions.

% Barrier certificates defined by k-degree polynomials
The partial derivative of the polynomials in Equation~\eqref{eq:poly} is required for ensuring the function meets Condition~\eqref{eq:complexconvex}. The partial derivative of the function is
\begin{equation}
    \label{eq:partial}
    \frac{\partial b}{\partial z_j}
    =
    \sum_{\boldalpha{} \in A_{n,k}}
    a_{\boldalpha{}} \alpha_j z_j^{ - 1}
    z^{\boldalpha{}}.
\end{equation}
We write
\begin{equation*}
    B(a,z) := b(a, z, \overline{z}) :=
    \sum_{\substack{
    (\boldalpha, \boldbeta) \in A_{2n, k} \\
    \boldalpha{} = (\alpha_1, \dots, \alpha_n)\\
    \boldbeta{} = (\alpha_{n+1}, \dots \alpha_{2n})
    }}
    a_{\boldalpha{},\boldbeta{}} z^{\boldalpha{}} \overline{z}^{\boldbeta{}},
\end{equation*}
where $a = (a_{\boldalpha,\boldbeta}) \in \mathbb{R}^{\abs{A_{2n,k}}}$ is a vector of real coefficients to be found and $\overline{z}^{\boldbeta{}} = \prod_{j=1}^n \overline{z_j}^{\alpha_{n+j}}$.

The following (polynomial) inequalities find the coefficient vector:
\begin{equation}
\label{eq:poly_ineq}
\begin{aligned}
    \textbf{find } & a^T \\
    \textbf{subject to }
    & B(a,z) \leq 0, \forall z \in \init{Z} \\
    & B(a,z) > 0, \forall z \in \unsafe{Z} \\
    & \dv{B(a,z)}{t} \leq 0, \forall z \in Z \\
    & B(a, z) \in \mathbb{R} \\
    & -1 \leq a_{\boldalpha{},\boldbeta{}} \leq 1.
\end{aligned}
\end{equation}
The coefficients, $a_{\boldalpha{},\boldbeta{}} \in \mathbb{R}$, are restricted to the range
$\begin{bmatrix}-1, 1\end{bmatrix}$ since any barrier certificate $B(a, z)$, can be normalised by dividing $B$ by the coefficient of greatest weight, $m = \max\abs{a_{\boldalpha{},\boldbeta{}}}$.
The resulting function $\frac{1}{m}B(a,z)$ is still a barrier certificate.
A barrier certificate generated from these polynomial inequalities can then freely be scaled up by multiplying it by a constant.

\subsection{An Algorithmic Solution}
% Methods of solving
One approach of solving the inequalities in \eqref{eq:poly_ineq} is to convert the system to real numbers and solve using sum of squares (SOS) optimisation \cite{Prajna2007};
another method is to use SMT solvers to find a satisfiable set of coefficients;
or it is possible to use neural network based approaches to find possible barriers \cite{peruffo2021automated,FOSSIL}.
% Our approach
We consider as a special case, an approach where $\dv{B(a,z)}{t} = 0$ rather than $\dv{B(a,z)}{t} \leq 0$, which allows the problem to be turned into a linear program.
This restriction allows us to consider a subset of barrier certificates that still ensures the safety of the system.
This is motivated by the fact that simple quantum systems of interest exhibit periodic behaviour; that is for all $t \in \mathbb{R}^+$, $z(t) = z(t + T)$ for some $T$.
The barrier must also exhibit periodic behaviour,\footnote{The barrier being periodic can be seen by interpreting the barrier as a function over time: $B(t) = B(z(t)) = B(z(t+T)) = B(t+T), \forall t \in \mathbb{R}^+$}
and this can be achieved by setting $\dv{B(a,z)}{t} = 0$.
Whilst there are other properties that ensure a function is periodic, these would involve non-polynomial terms such as trigonometric functions.
Further, linear programs tend to be solved faster than SOS methods.
This is because SOS programs are solved through semidefinite programming techniques, which are extensions of linear programs and therefore harder to solve.

% Transforming differential
We begin by transforming the differential constraint, $\dv{B(a,z)}{t} = 0$.
To obey the third condition for the complex-valued convex barrier certificate, we can substitute terms in Equation~\eqref{eq:complexdiff} with the partial derivatives from Equation~\eqref{eq:partial}.
Essentially one will end up with an equation of the form
\begin{equation*}
    (\mathbf{A} a)^\top \zeta = 0,
\end{equation*}
where $\zeta$ is a vector of all possible polynomial terms of $z_j, \overline{z_j}$ with degree less than $k$,\footnote{\eg for $k=2$ acceptable terms include $z_j^a, z_j z_l, z_j\overline{z_l}, \overline{z_j}^a, \overline{z_j}\overline{z_l}$ for $0 \leq a \leq 2$.
}
and $\mathbf{A}$ is a matrix of constant values.
By setting $\mathbf{A} a = \vec{0}$ the constraint is satisfied.
Therefore, each row of the resultant vector, $(\mathbf{A}a)_j = 0$, is added as a constraint to a linear program.

% Transforming real terms
To transform the real constraint ($B(a,z) \in \mathbb{R}$) note that if $x \in \mathbb{C}$, then $x \in \mathbb{R}$ if and only if $x = \overline{x}$.
Therefore, ${B(a, z) - \overline{B(a, z)} = 0}$ and we have
\begin{equation*}
\begin{aligned}
    B(a, z) - \overline{B(a, z)} & = 
    \sum_{\substack{
    (\alpha_j) \in A_{2n, k} \\
    \boldalpha{} = \{\alpha_1, \dots, \alpha_n\}\\
    \boldbeta = \{\alpha_{n+1}, \dots \alpha_{2n} \}
    }}
    a_{\boldalpha{},\boldbeta{}} z^{\boldalpha{}} \overline{z}^{\boldbeta{}}
    -
    \sum_{\substack{
    (\alpha_j) \in A_{2n, k} \\
    \boldalpha{}' = \{\alpha_1, \dots, \alpha_n\}\\
    \boldbeta' = \{\alpha_{n+1}, \dots \alpha_{2n} \}
    }}
    \overline{a}_{\boldalpha{}',\boldbeta'}  z^{\boldbeta'} \overline{z}^{\boldalpha{}'}
    \\
    & =
    \sum_{\substack{
    (\alpha_j) \in A_{2n, k} \\
    \boldalpha = \{\alpha_1, \dots, \alpha_n\}\\
    \boldbeta = \{\alpha_{n+1}, \dots \alpha_{2n} \}
    }}
    (a_{\boldalpha,\boldbeta} - \overline{a}_{\boldbeta,\boldalpha}) z^{\boldalpha} \overline{z}^{\boldbeta}.
\end{aligned}
\end{equation*}
The whole polynomial is equal to $0$ if all coefficients are $0$.
Thus, taking the coefficients and noting that $a_j$ are real gives the transformed constraints $a_{\boldalpha,\boldbeta} = a_{\boldbeta,\boldalpha} \text{ for } \boldalpha = (\alpha_j)_{j=1}^{n},\boldbeta=(\alpha_j)_{j=n+1}^{2n}, (\alpha_j) \in A_{2n,k}$.
These constraints to the coefficients are then also added to the linear program.

% Transforming unsafe and initial conditions
The final constraints we need to transform are the constraints on the initial and unsafe set: ${B(a,z) \leq 0}$ for $z \in \init{Z}$ and ${B(a,z) > 0}$ for $z \in \unsafe{Z}$, respectively.
We begin by noting that $B(a,z) = c + b(a, z, \overline{z})$ where $b(a,z,\overline{z})$ is a $k$-degree polynomial (with coefficients $a$) and $c \in \mathbb{R}$ is a constant.
When considering the differential and real constraint steps, $c$ is not involved in these equations since $c$ does not appear in the differential term and $c$ is cancelled out in the real constraint ($c - \overline{c} = c - c = 0$).

Considering the initial and unsafe constraints, we require that
\begin{equation*}
    \begin{aligned}
    \forall z \in \init{Z},\  c + b(a, z, \overline{z}) \leq 0, & \text{ and}
    \\
    \forall z \in \unsafe{Z},\  c + b(a, z, \overline{z}) > 0.
    \end{aligned}
\end{equation*}
Therefore, $c$ is bounded by
\begin{equation*}
    \max_{z \in \unsafe{Z}} -b(a, z, \overline{z}) < c \leq \min_{z \in \init{Z}} -b(a, z, \overline{z}).
\end{equation*}
Finding $c = \min_{z \in \init{Z}} -b(a, z, \overline{z})$ and then checking $\max_{z \in \unsafe{Z}} -b(a, z, \overline{z}) < c$ will ensure the initial and unsafe constraints are met for the barrier. The final computation is given in Algorithm~\ref{alg:computation}.

Note that the algorithm can fail since the function $b$ may divide the state space in such a way that a section of $\init{Z}$ may lie on the same contour as a section of $\unsafe{Z}$.
This means that either the function $b$ is unsuitable or the system is inherently unsafe.

\begin{algorithm}[t]
\caption{Computing the barrier certificate using linear programming}
\label{alg:computation}
\begin{algorithmic}[1]
    \State Solve the linear program
    \begin{align*}
        \textbf{find } & a^T \\
        \textbf{subject to }
        & \mathbf{A} a = \vec{0} & \\
        & a_{\boldalpha{},\boldbeta{}} = a_{\boldbeta{},\boldalpha{}} & \text{ for } \boldalpha{} = \{\alpha_j\}_{j=1}^{n},\boldbeta{}=\{\alpha_j\}_{j=n+1}^{2n}, \\
        & -1 \leq a_j \leq 1. & \text{ and }\{\alpha_j\}_{j=1}^{2n} \in A_{2n,k}
    \end{align*}
    
    \State $c \gets \min_{z \in \init{Z}} -b(a, z, \overline{z})$
    \State {\bf if}\, $c > \max_{z \in \unsafe{Z}} -b(a, z, \overline{z})$\, {\bf then\,  return} $B(a, z) = c + b(a, z, \overline{z})$
    \State {\bf else \, fail}
    \end{algorithmic}
\end{algorithm}

%% file: Sections/quantum.tex
We consider quantum systems that evolve within Hilbert spaces $\mathcal{H}^n = \mathbb{C}^{2^n}$ for $n \in \mathbb{N}$.
We use the computational basis states $\ket{j} \in \mathcal{H}^n$, for $0 \leq j < 2^n$, as an orthonormal basis within the space, where $(\ket{j})_l = \delta_{jl}$.\footnote{$\delta_{jl}$ is the Kronecker delta, which is 1 if $j=l$ and 0 otherwise.}
General quantum states, $\ket{\phi} \in \mathcal{H}^n$, can then be written in the form
\begin{equation*}
\ket{\phi} = \sum_{j=0}^{2^n -1} z_j \ket{j}, 
\end{equation*}
where $z_j \in \mathbb{C}$ and $\sum_{j=0}^{2^n - 1} \prob{z_j} = 1$.\footnote{For readers familiar with the Dirac notation, $z_j = \braket{j}{\phi}$ and $\overline{z_j} = \braket{\phi}{j}$.}
Quantum states reside within the unit circle of $\mathbb{C}^{2^n}$.
For simplicity, we consider quantum systems that evolve according to the \Schrodinger{} equation
\begin{equation*}
\frac{\dd{\ket{\phi}}}{\dd{t}} = - \iu \hat{H} \ket{\phi},
\end{equation*}
where $\hat{H}$ is a Hamiltonian, a complex matrix such that $\hat{H} = \hat{H}^\dagger = \overline{\hat{H}^\top}$; and $\ket{\phi}$ is a quantum state.\footnote{We set the Planck constant $\hbar = 1$ in the \Schrodinger{} equation.}
In the rest of this section, we make use of Algorithm~\ref{alg:computation} in order to find suitable barrier certificates for operations that are commonly used in quantum computers.

\subsection{Hadamard Operation Example}
\label{sec:application:had}
The evolution of the Hadamard operation, $ H = \frac{1}{\sqrt{2}}\begin{pmatrix} 1 & 1 \\ 1 & -1 \end{pmatrix} $, is given by $ \hat{H}_{H} = \begin{pmatrix} 1 & 1 \\ 1 & -1 \end{pmatrix} $ and $\ket{\phi}$ is one qubit, $z_0 \ket{0} + z_1 \ket{1}$.
We have $z(t) = \begin{pmatrix}z_0(t)\\ z_1(t)\end{pmatrix}$ and 
\begin{equation*}
\label{eq:had_ham}
\dot{z} = -\iu\hat{H}_{H}z =  -\iu\begin{pmatrix}z_0 + z_1\\ z_0 - z_1 \end{pmatrix}.
\end{equation*}
The system evolves over the surface of the unit sphere, $Z = \{(z_0, z_1) \in \mathbb{C}^2 : \prob{z_0} + \prob{z_1} = 1\}$.
The initial set is defined as $\init{Z} = \{(z_0, z_1) \in Z : \prob{z_0} \geq 0.9 \}$ and the unsafe set as $\unsafe{Z} = \{(z_0, z_1) \in Z : \prob{z_0} \leq 0.1 \}$.
Note that the definitions of $\init{Z}$ and $\unsafe{Z}$ are restricted by $Z$, therefore $\prob{z_1} \leq 0.1$ and $\prob{z_1} \geq 0.9$ for $\init{Z}$ and $\unsafe{Z}$ respectively.
A barrier function computed by our Algorithm~\ref{alg:computation} is
\begin{equation*}
B(z) = \frac{11}{5} - 3z_0\overline{z_0} - z_0\overline{z_1} - \overline{z_0}z_1 - z_1\overline{z_1}.
% = \frac{6}{5} - 2 z_0 \overline{z_0} - z_0\overline{z_1} - \overline{z_0} z_1
\end{equation*}
By rearranging and using properties of the complex conjugate, we find that
\begin{equation*}
B(z) = 2 (\frac{1}{10} - \prob{z_0} + \frac{1}{2} - \Re{z_0\overline{z_1}}).
\end{equation*}
The derivation is given in Appendix~\ref{app:deriv}.
The first term of the barrier ($\frac{1}{10} - \prob{z_0}$) acts as a restriction on how close to $\ket{0}$ as $\ket{\phi}$ evolves, whereas the second term ($\frac{1}{2} - \Re{z_0\overline{z_1}}$) is a restriction on the phase of the quantum state.
Next, we double check that $B$ is indeed a barrier certificate.

\begin{figure}[t]
    \centering
    \begin{subfigure}[t]{.4\textwidth}
        \centering
        \includegraphics[width=\textwidth]{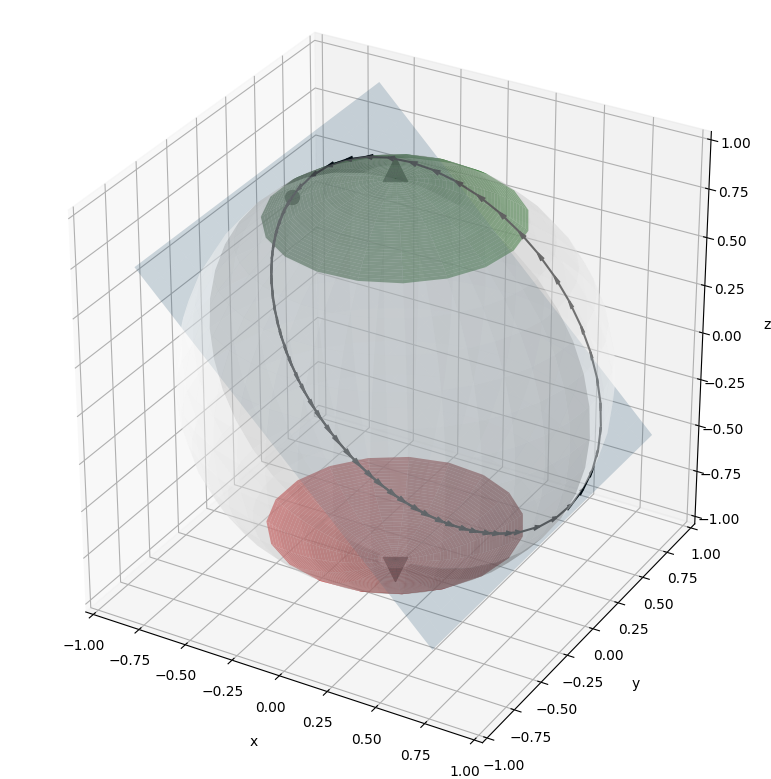}
        \caption{Isometric view of system}
        \label{fig:exH1}
    \end{subfigure}
     \hfill
     \begin{subfigure}[t]{.45\textwidth}
         \centering
         \includegraphics[width=\textwidth]{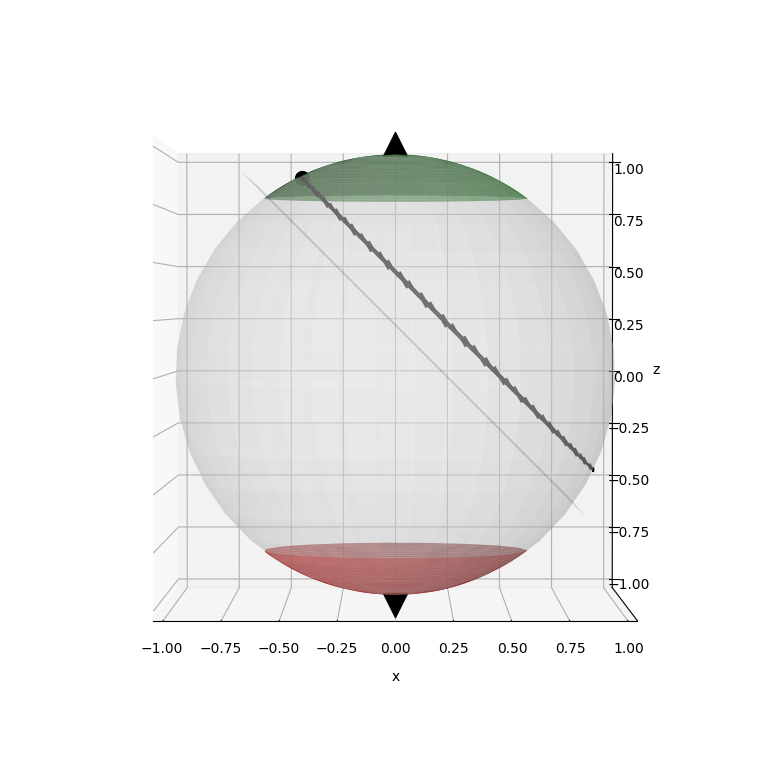}
         \caption{Top down view of system}
         \label{fig:exH2}
     \end{subfigure}
    \caption{System evolution on a Bloch sphere.
        The initial state of the system is $\sqrt{0.9}\ket{0} + \iu\sqrt{0.1}\ket{1}$ (the black dot) and evolves according to the black line (in an anti-clockwise rotation with a period of $t = \pi$).
        The green surface around the north pole ($\ket{0}$) is the initial region, $\init{Z}$, and the red surface around the south pole ($\ket{1}$) is the unsafe region, $\unsafe{Z}$.
        The blue surface is the plane of the barrier function when $B(z) = 0$, with $x < -z$ being the unsafe region.}
        \label{fig:exH}
\end{figure}

\begin{proposition}
\label{prop:had_safe}
The system evolving according to Equation~\eqref{eq:had_ham}, initial set $Z_0$ and unsafe set $Z_u$ is safe.
\end{proposition}
The proposition is proved in Appendix~\ref{app:had_safe}.
A visualisation on a Bloch sphere representation of the example system and its associate barrier are given in Figure~\ref{fig:exH}.

\subsection{Phase Operation Example}
\begin{figure}[t]
    \centering
    \begin{subfigure}[t]{.45\textwidth}
\centering
        \includegraphics[width=\textwidth]{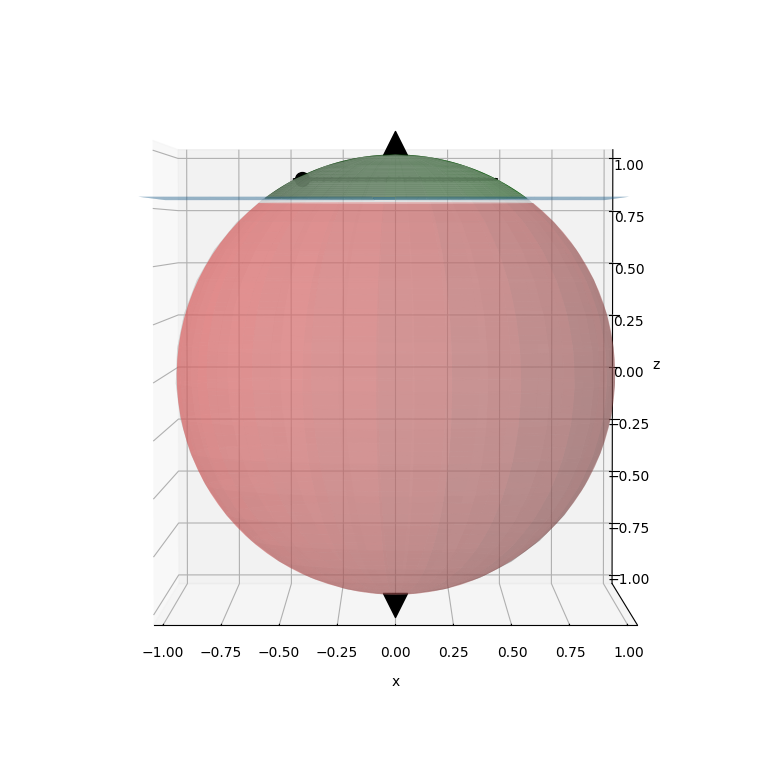}
        \caption{Evolution with initial and unsafe states $Z^1$. The barrier at $B_1(z) = 0$ is a flat plane that borders $Z_0^1$.}
        \label{fig:exS1}
    \end{subfigure}
     \hfill
     \begin{subfigure}[t]{.45\textwidth}
         \centering
         \includegraphics[width=\textwidth]{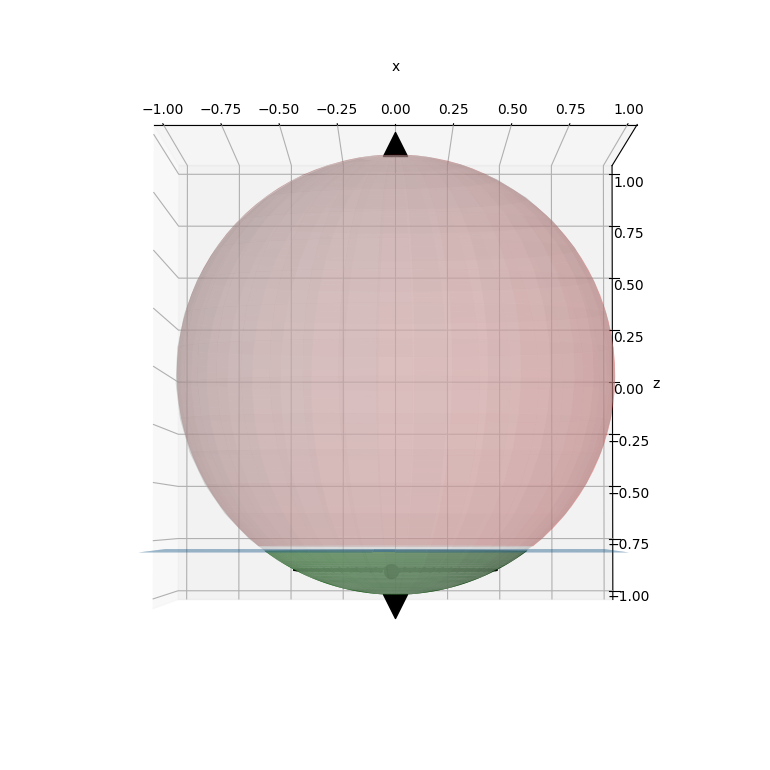}
         \caption{Evolution with initial and unsafe states $Z^2$. Similarly, $B_2(z) = 0$ is a flat plane that borders $Z_0^2$.}
         \label{fig:exS2}
     \end{subfigure}
    \caption{State evolution of \eqref{eq:S_ham} demonstrated on a Bloch sphere.}
    \label{fig:exSgate}
\end{figure}

The evolution of the phase operation
$S = \begin{pmatrix} 1 & 0 \\ 0 & \iu \end{pmatrix}$
is given by the Hamiltonian
$\hat{H}_{S} = \begin{pmatrix} 1 & 0 \\ 0 & -1 \end{pmatrix}$ for a single qubit $z_0 \ket{0} + z_1 \ket{1}$.
Thus, the evolution of the system for $z(t) = \begin{pmatrix} z_0(t) \\ z_1(t) \end{pmatrix}$ is 
\begin{equation}
\label{eq:S_ham}
\dot{z} = -\iu \begin{pmatrix} z_0 \\ -z_1 \end{pmatrix}.
\end{equation}
Again, $Z$ represents the unit sphere as described previously.
Two pairs of safe and unsafe regions are given. The first pair $Z_1 = (\init{Z^1}, \unsafe{Z^1})$ is given by
\begin{equation*}
\begin{aligned}
\init{Z^1} = \{ (z_0, z_1) \in Z : \prob{z_0} \geq 0.9 \},
& &
\unsafe{Z^1} = \{(z_0, z_1) \in Z : \prob{z_1} > 0.11 \};
\end{aligned}
\end{equation*}
and the second pair $Z_2 = (\init{Z^2}, \unsafe{Z^2})$ is given by
\begin{equation*}
\begin{aligned}
\init{Z^2} = \{ (z_0, z_1) \in Z : \prob{z_1} \geq 0.9 \},
& &
\unsafe{Z^2} = \{(z_0, z_1) \in Z : \prob{z_0} > 0.11 \}.
\end{aligned}
\end{equation*}
The pair $Z^1$ starts with a system that is close to the $\ket{0}$ state and ensures that the system cannot evolve towards the $\ket{1}$ state.
The pair $Z^2$ has similar behaviour with respective states $\ket{1}$ and $\ket{0}$.
The system for each pair of constraints is considered safe by  the following barriers computed by 
Algorithm~\ref{alg:computation}:
\begin{equation*}
    \begin{aligned}
        B_1(z) = 0.9 - z_0 \overline{z_0},
        & &
        B_2(z) = 0.9 - z_1 \overline{z_1},
    \end{aligned}
\end{equation*}
where $B_1$ is the barrier for $Z^1$ and $B_2$ is the barrier for $Z^2$.\footnote{These barriers can similarly be written using the Dirac notation.}
The system with different pairs of regions can be seen on Bloch spheres in Figure~\ref{fig:exSgate}.
Again, both functions $B_1$ and $B_2$ are valid barrier certificates.

\begin{proposition}
    The system given by Equation~\ref{eq:S_ham} with the set of initial states $\init{Z^1}$ and the unsafe set $\unsafe{Z^1}$ is safe.
\end{proposition}

\begin{proposition}
    The system given by Equation~\ref{eq:S_ham} with the set of initial states $\init{Z^2}$ and the unsafe set $\unsafe{Z^2}$ is safe.
\end{proposition}

The proofs are omitted as they are similar to the proof given in Proposition~\ref{prop:had_safe}.
These barriers give bounds on how the system evolves, \ie the system must only change the phase of the system and not the amplitude.
This can be applied in general by combining barriers to show how a (disturbed) system is restricted in its evolution.

\subsection{Controlled-NOT Operation Example}
The final example we consider is the controlled-NOT (CNOT) operation acting on two qubits; a control qubit, $\ket{\phi_c}$, and a target qubit, $\ket{\phi_t}$, with the full quantum state being $\ket{\phi_c \phi_t}$.
The CNOT operation performs the NOT operation on a target qubit ($\ket{0} \to \ket{1}$ and $\ket{1} \to \ket{0}$) if the control qubit is set to $\ket{1}$ and does nothing if the control qubit is set to $\ket{0}$.
The CNOT operation and its associated Hamiltonian are given by
\begin{equation*}
\begin{aligned}
    \text{CNOT} = \begin{pmatrix}
        1 & 0 & 0 & 0 \\
        0 & 1 & 0 & 0 \\
        0 & 0 & 0 & 1 \\
        0 & 0 & 1 & 0
    \end{pmatrix}
    &, &
    \hat{H}_{\text{CNOT}} = \begin{pmatrix}
        0 & 0 & 0 & 0 \\
        0 & 0 & 0 & 0 \\
        0 & 0 & 1 & -1 \\
        0 & 0 & -1 & 1
    \end{pmatrix}.
\end{aligned}
\end{equation*}
The system $z(t) = (z_j(t))_{j=0, \dots, 3}$ evolves according to
\begin{equation*}
    \dot{z} = -\iu\begin{pmatrix}
        0 \\
        0 \\
        z_2 - z_3 \\
        -z_2 + z_3
    \end{pmatrix}.
\end{equation*}
This system evolves over $Z = \{(z_0, \dots, z_3) \in \mathbb{C}^4 : \sum_{j=0}^3 |z_j|^2 = 1\}$.
Using this as our system, various initial and unsafe regions can be set up to reason about the behaviour of the CNOT operation.

\subsubsection{Control in $\ket{0}$}
Here we consider the following initial and unsafe regions
\begin{equation*}
\begin{aligned}
    & \init{Z} = \{ (z_j)_{j=0}^{3} \in \mathbb{C}^4 : \prob{z_0} \geq 0.9 \}, \\
    & \unsafe{Z} = \{ (z_j)_{j=0}^{3} \in \mathbb{C}^4 : \prob{z_1} + \prob{z_2} + \prob{z_3} \geq 0.11 \}.
\end{aligned}
\end{equation*}
The initial set, $\init{Z}$, encapsulates the quantum states that start in the $\ket{00}$ state with high probability and $\unsafe{Z}$ captures the states that are not in the initial region with probability greater than $0.11$.
These regions capture the behaviour that the quantum state should not change much when the control qubit is in the $\ket{0}$ state.
Using Algorithm~\ref{alg:computation}, the barrier $B(z) = 0.9 - z_0 \overline{z_0}$ can be generated to show that the system is safe.

A similar example can be considered where the initial state $\ket{00}$ is replaced with $\ket{01}$ instead (swap $z_0$ and $z_1$ in $\init{Z}$ and $\unsafe{Z}$).
The behaviour that the state of the system should not change much is still desired; the function $B(z) = 0.9 - z_1 \overline{z_1}$ is computed as a barrier to show this behaviour is met.

\subsubsection{Control in $\ket{1}$}
Now consider when the initial region has the control qubit near the state $\ket{1}$.
The following regions are considered:
\begin{equation*}
\begin{aligned}
    & \init{Z} = \{ (z_j)_{j=0}^{3} \in \mathbb{C}^4 : \prob{z_2} \geq 0.9 \}, \\
    & \unsafe{Z} = \{ (z_j)_{j=0}^{3} \in \mathbb{C}^4 : \prob{z_1} + \prob{z_2} \geq 0.11 \}.
\end{aligned}
\end{equation*}
This system starts close to the $\ket{10}$ state and the evolution should do nothing to the control qubit.
Note that the specified behaviour does not captures the NOT behaviour on the target qubit.
Our Algorithm~\ref{alg:computation} considers this system safe by outputting the barrier certificate $B(z) = 0.9 - z_2\overline{z_2} - z_3\overline{z_3}$.
This is also the barrier if the system were to start in the $\ket{11}$ state instead.

%% file: Sections/appendix.tex
\section{Proof of Theorem~\ref{thm:complex_convex_safe}}
\label{app:complex_convex_safe}
The proof is similar to the intuition given for Theorem~\ref{thm:real_convex_safe}.

Assume by contradiction that the system has a complex-valued convex barrier certificate, but the system is not safe.
Therefore, there is an initial state $z(0) \in \init{Z}$ and time $T \in \mathbb{R}^{+}$ such that $z(T) \in \unsafe{Z}$.
By the definition of our convex barrier certificate, we have that $B(z(0)) \leq 0$ and $B(z(T)) > 0$.
Thus, the barrier must grow positively at some point during the system evolution.
However, we have that $\dv{B(z(t))}{t} \leq 0$ for all $t \in \mathbb{R}^+$ based on Equation~\eqref{eq:complexconvex}.
The system cannot grow positively and so we have a contradiction.
Therefore, the system must be safe.
\qed

\section{Proof of Proposition~\ref{prop:complexisconvex}}
\label{app:complexisconvex}
Let $\dot{z} = f(z)$ be a system over $Z$ with $Z_0$ and $Z_u$ being the initial and unsafe sets as before.
Let $\mathcal{B}$ denote the set of (complex-valued convex) barrier certificates such that for any $B \in \mathcal{B}$ the system $f(z)$ is safe.
Take $B_1, B_2 \in \mathcal{B}$ and consider the function $B(z) = \lambda B_1(z) + (1-\lambda) B_2(z)$, where $\lambda \in [0,1]$.
Since $B_1(z) \leq 0$ and $B_2(z) \leq 0$ for all $z \in Z_0$, then $B(z) \leq 0$ as well.
A similar argument holds for $B(z) > 0$ for all $z \in Z_u$.
Finally, consider the differential equation $\dBdt{}$. It is trivial to see that
\begin{equation*}
    \dBdt = \lambda {\frac{\dd B_1}{\dd t}} + (1-\lambda) {\frac{\dd B_2}{\dd t}} \leq 0,
\end{equation*}
because differentiation is linear; and ${\frac{\dd B_1}{\dd t}}, {\frac{\dd B_2}{\dd t}} \leq 0$ for all $z \in Z$.
Therefore, $B$ satisfies the properties of a barrier certificate for $f(z)$ and so $B \in \mathcal{B}$.
Hence, $\mathcal{B}$ is convex.
\qed

\section{Derivation of Barrier for Hadamard System}
\label{app:deriv}
By substituting $z_j \overline{z_j} = \abs{z_j}^2$ and noting that $\Re{z} = z + \overline{z}$ for any $z \in \mathbb{C}$, we have that
\begin{equation*}
    B(z) = \frac{11}{5} - 3\prob{z_0} - \Re{z_0\overline{z_1}} - \prob{z_1}.
\end{equation*}

Since ${\prob{z_1} = 1 - \prob{z_0}}$ (due to properties of quantum systems), we then have
\begin{equation*}
    B(z) = \frac{6}{5} - 2\prob{z_0} - \Re{z_0\overline{z_1}},
\end{equation*}

and by simply rearranging we get

\begin{equation*}
    B(z) = 2 (\frac{1}{10} - \prob{z_0} + \frac{1}{2} - \Re{z_0\overline{z_1}}).
\end{equation*}

\section{Proof of Proposition~\ref{prop:had_safe}}
\label{app:had_safe}
We prove this by showing that $B$ meets the conditions of a convex barrier certificate (given in Definition~\ref{def:complexconvex}). Safety is then guaranteed from Theorem~\ref{thm:complex_convex_safe}.

Firstly, consider $z \in \init{Z}$. As $\prob{z_0} \geq 0.9$, then $B(z) \leq 2(-\frac{4}{5} - \Re{z_0\overline{z_1}})$.
Further, it can be seen that
\begin{equation*}
\abs{\Re{z_0\overline{z_1}}} = \abs{\Re{z_{0}} \Re{z_{1}} + \Im{z_{0}} \Im{z_{1}}} < 1 \times \sqrt{\frac{1}{10}} + 1 \times \sqrt{\frac{1}{10}} = \sqrt{\frac{2}{5}}.
\end{equation*}
Note that we are taking the maximal possible value of each component and therefore this is larger than the maximal value of $\Re{z_0 \overline{z_1}}$.
Thus, 
\begin{equation*}
{B(z) \leq 2(-\frac{4}{5} - \Re{z_0\overline{z_1}}) < 2(-\frac{4}{5} + \sqrt{\frac{2}{5}}) < 0}.
\end{equation*}
A similar argument can be made for when $z \in \unsafe{Z}$ and it can be shown that $B(z) > 0$.
Finally, we use Equations~\eqref{eq:complexdiff} and \eqref{eq:had_ham} to get
\begin{equation*}
    \begin{aligned}
    \dBdt{} & = - \iu
    \Big(- (2\overline{z_0} + \overline{z_1})(z_0 + z_1) 
    - (\overline{z_0}) (z_0 - z_1) 
    \\ 
    &
    + (2z_0 + z_1) (\overline{z_0} + \overline{z_1}) 
    + (z_0) (\overline{z_0} - \overline{z_1}) \Big) \\
    & = -\iu
    \Big( 
    -2 \overline{z_0} z_1 - z_0 \overline{z_1} + \overline{z_0} z_1
    + 2 z_0 \overline{z_1} + \overline{z_0} z_1 - z_0 \overline{z_1}
    \Big) \\
    & = 0 , \forall z \in Z.
    \end{aligned}
\end{equation*}
Therefore, the system meets the conditions given in Equations \eqref{eq:complexinit}, \eqref{eq:complexunsafe} and \eqref{eq:complexconvex}; the system is safe.
\qed

%% file: main.bbl
\begin{thebibliography}{10}
\providecommand{\url}[1]{\texttt{#1}}
\providecommand{\urlprefix}{URL }
\providecommand{\doi}[1]{https://doi.org/#1}

\bibitem{FOSSIL}
Abate, A., et~al.: {FOSSIL}: A software tool for the formal synthesis of
  {Lyapunov} functions and barrier certificates using neural networks. In:
  Proceedings of the 24th International Conference on Hybrid Systems:
  Computation and Control. {ACM} (2021). \doi{10.1145/3447928.3456646}

\bibitem{ames2019control}
Ames, A.D., et~al.: Control barrier functions: Theory and applications. In:
  18th European control conference (ECC). pp. 3420--3431. IEEE (2019)

\bibitem{bak2018t}
Bak, S.: {t-B}arrier certificates: A continuous analogy to {k}-induction. In:
  6th IFAC Conference on Analysis and Design of Hybrid Systems. pp. 145--150
  (2018). \doi{https://doi.org/10.1016/j.ifacol.2018.08.025}

\bibitem{Burgholzer2021}
Burgholzer, L., Wille, R.: Advanced equivalence checking for quantum circuits.
  IEEE Transactions on Computer-Aided Design of Integrated Circuits and Systems
   \textbf{40},  1810--1824 (2021). \doi{10.1109/TCAD.2020.3032630}

\bibitem{QBricks}
Chareton, C., et~al.: An automated deductive verification framework for
  circuit-building quantum programs. In: Programming Languages and Systems. pp.
  148--177. Springer International Publishing (2021).
  \doi{10.1007/978-3-030-72019-3\_6}

\bibitem{modelchecking}
Clarke, E.M., et~al.: Model checking, 2nd Edition. {MIT} Press (2018)

\bibitem{Cousot2001}
Cousot, P.: Abstract Interpretation Based Formal Methods and Future Challenges,
  pp. 138--156. Springer Berlin Heidelberg (2001).
  \doi{10.1007/3-540-44577-3\_10}

\bibitem{Cousot77}
Cousot, P., Cousot, R.: Abstract interpretation: A unified lattice model for
  static analysis of programs by construction or approximation of fixpoints.
  In: Proceedings of the 4th ACM SIGACT-SIGPLAN Symposium on Principles of
  Programming Languages. p. 238–252 (1977). \doi{10.1145/512950.512973}

\bibitem{Fang2013}
Fang, T., Sun, J.: Stability analysis of complex-valued nonlinear differential
  system. Journal of Applied Mathematics  \textbf{2013},  621957 (2013).
  \doi{10.1155/2013/621957}

\bibitem{Franzle19}
Fr\"{a}nzle, M., Chen, M., Kr\"{o}ger, P.: In memory of {Oded Maler}: Automatic
  reachability analysis of hybrid-state automata. ACM SIGLOG News
  \textbf{6}(1),  19–39 (2019). \doi{10.1145/3313909.3313913}

\bibitem{SQIR}
Hietala, K., et~al.: Proving quantum programs correct. In: 12th International
  Conference on Interactive Theorem Proving. pp. 21:1--21:19. Leibniz
  International Proceedings in Informatics (LIPIcs) (2021).
  \doi{10.4230/LIPIcs.ITP.2021.21}

\bibitem{Honarvar2020}
Honarvar, S., Mousavi, M.R., Nagarajan, R.: Property-based testing of quantum
  programs in {Q\#}. In: Proceedings of the IEEE/ACM 42nd International
  Conference on Software Engineering Workshops. pp. 430--435 (2020).
  \doi{10.1145/3387940.3391459}

\bibitem{Jagtap2020}
Jagtap, P., Soudjani, S., Zamani, M.: Formal synthesis of stochastic systems
  via control barrier certificates. IEEE Transactions on Automatic Control
  \textbf{66}(7),  3097--3110 (2021). \doi{10.1109/TAC.2020.3013916}

\bibitem{lavaei2021automated}
Lavaei, A., Soudjani, S., Abate, A., Zamani, M.: Automated verification and
  synthesis of stochastic hybrid systems: A survey. arXiv preprint
  arXiv:2101.07491  (2021)

\bibitem{QHL}
Liu, J., et~al.: Formal verification of quantum algorithms using quantum
  {H}oare logic. Lecture Notes in Computer Science  \textbf{11562 LNCS},
  187--207 (2019). \doi{10.1007/978-3-030-25543-5\_12}

\bibitem{Mitchell07}
Mitchell, I.M.: Comparing forward and backward reachability as tools for safety
  analysis. In: Proceedings of the 10th International Conference on Hybrid
  Systems: Computation and Control. p. 428–443. Springer-Verlag (2007)

\bibitem{peruffo2021automated}
Peruffo, A., Ahmed, D., Abate, A.: Automated and formal synthesis of neural
  barrier certificates for dynamical models. In: International Conference on
  Tools and Algorithms for the Construction and Analysis of Systems. pp.
  370--388. Springer (2021). \doi{10.1007/978-3-030-72016-2\_20}

\bibitem{Prajna2007}
Prajna, S., Jadbabaie, A., Pappas, G.J.: A framework for worst-case and
  stochastic safety verification using barrier certificates. IEEE Transactions
  on Automatic Control  \textbf{52},  1415--1428 (2007).
  \doi{10.1109/TAC.2007.902736}

\bibitem{FVBook}
Seligman, E., Schubert, T., Kumar, M.V.A.K.: Formal Verification: An Essential
  Toolkit for Modern {VLSI} Design. Morgan Kaufmann Publishers Inc. (2015)

\bibitem{SS2014precise}
Soudjani, S., Abate, A.: Precise approximations of the probability distribution
  of a {M}arkov process in time: an application to probabilistic invariance.
  In: International Conference on Tools and Algorithms for the Construction and
  Analysis of Systems. pp. 547--561. Springer (2014).
  \doi{10.1007/978-3-642-54862-8\_45}

\bibitem{SS2015quantitative}
Soudjani, S., Abate, A.: Quantitative approximation of the probability
  distribution of a {M}arkov process by formal abstractions. {Logical Methods
  in Computer Science}  \textbf{11} (2015). \doi{10.2168/LMCS-11(3:8)2015}

\bibitem{Giallar}
Tao, R., et~al.: Giallar: Push-button verification for the {Q}iskit quantum
  compiler. In: Proceedings of the 43rd ACM SIGPLAN International Conference on
  Programming Language Design and Implementation. p. 641–656 (2022).
  \doi{10.1145/3519939.3523431}

\bibitem{ZXcalc}
van~de Wetering, J.: {ZX}-calculus for the working quantum computer scientist.
  arXiv preprint arXiv:2012.13966  (2020)

\bibitem{Wisniewski13}
Wisniewski, R., Sloth, C.: Converse barrier certificate theorem. In: 52nd IEEE
  Conference on Decision and Control. pp. 4713--4718 (2013).
  \doi{10.1109/CDC.2013.6760627}

\bibitem{Yu21}
Yu, N., Palsberg, J.: Quantum abstract interpretation. In: Proceedings of the
  42nd ACM SIGPLAN International Conference on Programming Language Design and
  Implementation. p. 542–558 (2021). \doi{10.1145/3453483.3454061}

\bibitem{Zulehner2017}
Zulehner, A., Wille, R.: Advanced simulation of quantum computations. IEEE
  Transactions on Computer-Aided Design of Integrated Circuits and Systems
  \textbf{38},  848--859 (2017). \doi{10.1109/TCAD.2018.2834427}

\end{thebibliography}
